\documentclass{kluwer}
\usepackage[dvips]{graphicx}

\newcommand{\SN}{{_{\rm SN}}}

\newcommand{\SNII}{{_{\rm SN\,II}}}

\newcommand\la{\;
  \raise0.3ex\hbox{$<$\kern-0.75em\raise-1.1ex\hbox{$\sim$
  }}\;\hskip-2pt }
\newcommand\ga{\;
  \raise0.3ex\hbox{$>$\kern-0.75em\raise-1.1ex\hbox{$\sim$
  }}\;\hskip-2pt }

\newcommand{\cmcube}{\,{\rm cm^{-3}}}
\newcommand{\cmcb}{\,{\rm cm^{3}}}

\newcommand{\erg}{\,{\rm erg}}
\newcommand{\gm}{\,{\rm g}}

\newcommand{\K}{\,{\rm K}}
\newcommand{\kpc}{\,{\rm kpc}}
\newcommand{\pc}{\,{\rm pc}}

\newcommand{\muG}{\,\mu{\rm G}}

\newcommand{\s}{\,{\rm s}}
\newcommand{\yr}{\,{\rm yr}}
\newcommand{\Myr}{\,{\rm Myr}}
\newcommand{\yana}[3]{, Astron. Astrophys. {\bf #2}, #3 (#1).}

\newcommand{\pass}[1]{, Astrophys. Spa. Sci., in press (#1).}

\newcommand{\ymn}[3]{, Monthly Notices Roy. Astron. Soc. {\bf #2}, #3 (#1).}

\newcommand{\yaj}[3]{, Astronom. J. {\bf #2}, #3 (#1).}
\newcommand{\yapj}[3]{, Astrophys. J. {\bf #2}, #3 (#1).}

\newcommand{\yapjl}[3]{, Astrophys. J. {\bf #2}, #3 (#1).}

\newcommand{\yprep}[2]{, (preprint).}

\newcommand{\beq}{\begin{equation}}
\newcommand{\eeq}{\end{equation}}
\newcommand{\bea}{\begin{eqnarray}}
\newcommand{\eea}{\end{eqnarray}}
\begin{document}
\begin{article}
\begin{opening}


\title{Self-regulating supernova heating in interstellar medium simulations}
\author{Graeme R.\ \surname{Sarson} \& Anvar \surname{Shukurov}}
\institute{School of Mathematics \& Statistics, University of
Newcastle, Newcastle NE1 7RU, U.K.}
\author{{\AA}ke \surname{Nordlund}}
\institute{NBIfAFG \& TAC,
Juliane Maries Vej 30, DK-2100 Copenhagen \O, Denmark}
\author{Boris \surname{Gudiksen}}
\institute{Inst.\ Solar Physics, Royal Swedish Acad.\ of Sciences,
SE-106 91 Stockholm, Sweden}
\author{Axel \surname{Brandenburg}}
\institute{NORDITA, Blegdamsvej 17, DK-2100 Copenhagen \O, Denmark}

\runningauthor{G. R.\ Sarson et al.}
\runningtitle{Self-regulating SNe in ISM simulations}
\date{\today}


\begin{abstract}
Numerical simulations of the multi-phase interstellar medium have been carried out,
using a 3D, nonlinear, magnetohydrodynamic, shearing-box model,
with random motions driven by supernova explosions.
These calculations incorporate the effects of magnetic fields 
and rotation in 3D;
these play important dynamical roles in the galaxy,
but are neglected in many other simulations.
The supernovae driving the motions are not arbitrarily imposed,
but occur where gas accumulates into cold, dense clouds;
their implementation uses a physically motivated model
for the evolution of such clouds.  
The process is self-regulating,
and produces mean supernova rates as part of the solution.
Simulations with differing mean density show a power law relation
between the supernova rate and density, 
with exponent 1.7;
this value is within the range suggested from observations
(taking star formation rate as a proxy for supernova rate).
The global structure of the supernova driven medium
is strongly affected by the presence of magnetic fields;
e.g.\ for one solution the filling factor of hot gas is found to vary from
0.19 (with no field) to 0.12 (with initial mid-plane field $B_{0}=6\muG$).
\end{abstract}
\keywords{ISM, turbulence, magnetic fields, supernovae}

\end{opening}


\section{Introduction}

The dynamical state of the interstellar medium (ISM)
remains relatively poorly understood,
especially in the presence of magnetic fields.
Whilst the complexity of the topic still precludes
a single all-encompass\-ing model,
there are many aspects of the problem that can now be addressed
by well-designed numerical simulations.
Recent 3D models of relevance include those of Korpi et al.\ (1999a,b),
and of de Avillez (2000), de Avillez \& Mac Low (2002).
These models focus on small Cartesian boxes,
with typical side-lengths of order $1 \kpc$,
rather that addressing the full galactic disc.
For similar reasons of computational resolution,
none attempt to model all of the phases of gas present,
from the hottest ionized plasma
(with density $\rho \sim 10^{-27}\gm/\-\cmcb$, temperature $T \sim 10^6\K)$
to the cold molecular clouds
($\rho > 10^{-21}\gm/\-\cmcb$, $T < 10\K)$.
And each of the models neglects or simplifies some aspects of the
physics of the problem;
at this early stage of study, that is to be expected.

\section{Method}

The current model ---
a development of the earlier work of Korpi et al.\ (1999a,b),
already partially reported in Shukurov et al.\ (2003) ---
focuses in particular on the dynamics of the hot ISM gas,
under the interaction of supernova (SN) driving
with classical magnetohydrodynamic (MHD) forces.
We model a relatively small box (of horizontal cross-section $(0.25 \kpc)^{2}$),
symmetric about the galactic mid-plane ($z=0$);
this is nevertheless large enough to follow the evolution
of turbulent ISM structures on observable scales.
We neglect the very coldest gas,
focusing on the warm, hot and ionized phases ($T \gtrsim 500$ K).
This choice retains enough cold gas
to fit a major fraction of the mass into a small fraction of the volume
--- allowing interesting 3D structures to form ---
but allows us to run with a relatively coarse resolution of $4 \pc$.
Consistent with this restriction,
we do not incorporate self-gravity into our solution,
instead assuming the fixed vertical gravity profile 
from Ferri\`ere (1998).
Beyond the fine details,
the numerical model is essentially as described in
Brandenburg et al.\ (1995). The treatment of the thermodynamics, 
adapted for the ISM problem,
is largely as in Korpi et al.\ (1999a,b);
this approach is indebted to earlier work in 2D
(e.g. Passot, V\'{a}squez-Semadeni \& Pouquet, 1995; 
Gazol-Pati\~{n}o \& Passot, 1999).

The radiative cooling function, $\Lambda$,
is that given by Rosen, Bregman \& Norman (1993)
(after that of Chiang \& Bregman, 1998),
truncated at $500\K$ 
to avoid the gas evolving towards the colder phases we wish to neglect.
Heating by UV absorption, $\Gamma_{\rm UV}$,
is imposed with a net heating rate of $0.05\erg \gm^{-1} \s^{-1}$ 
for temperatures $T \lesssim 10,000\K$
(e.g. Wolfire et al., 1995).
This is not a major energy input,
but again helps to prevent gas from cooling and contracting
into phases we do not aim to resolve.
We assume a fixed fraction of Helium of 8\% (by number),
and a fixed ionization fraction of 0.7,
in implementing these two parameterisations.

The principal driving is the supernova heating $\Gamma\SN$,
which is implemented by the occasional, instantaneous, input
of $E\SN=10^{51}\erg$.
This energy is locally distributed with a normalised profile of form 
$\exp(-(r/d)^6)$,
at distance $r$ about the SN location;
this profile, with the width taken as $d=16\pc$,
permits adequate numerical resolution.
The internal energy then evolves naturally;
some remains in thermal form or is lost to radiative cooling,
some is transformed to kinetic and magnetic energy
by normal gas dynamic and MHD processes.
We find this scheme works well, with one refinement.
In gas above a certain density,
injecting $E\SN$ in this way
distributes the energy over a significant total mass;
the central gas is then insufficiently heated,
compared with the point-source energy injection we are trying to model.
The resultant state undergoes rapid radiative cooling,
losing too much energy this way,
and converting an unrealistically small amount 
to kinetic energy by pressure-driven expansion.
To avoid this problem, we model the initial phase of SN expansion
by creating a cavity of width $d$ of low density gas,
and moving the original mass to a symmetric shell beyond this.
Tests on the subsequent evolution show that Sedov-type expansion
is well replicated,
and both energy and mass are well conserved.
With the SN heating implemented in this way,
about 15\% of the injected energy
is typically transformed to kinetic energy.

We determine SN frequency and location
using the following scheme from Gudiksen (1999).
(Although again, the method has antecedents:
e.g.\ Bania \& Lyon, 1980;
Passot et al., 1995.)
We identify regions with
$\rho > 10^{-24}\gm/\-\cmcb$, $T < 4000\K$
as clouds,
and only permit SN~II within such clouds.
The mass of gas in such regions, $M_{\rm c}$,
is monitored, and the instantaneous SN~II frequency,
$\nu\SNII$, is taken as
\[
\nu\SNII=\frac{M_{\rm c} X_{*} X\SN}{M\SN\tau_{\rm c}}\;.
\]
Here $X_*=0.02$ is the gas mass fraction converted into stars
(cf.\ Wilson \& Matthews, 1995);
$X\SN=0.1$ is the fraction of the stellar mass in SN progenitors,
obtained from a suitable initial mass function
(e.g. Padoan et al., 1997);
$M\SN=10\,M_\odot$ is the SN progenitor mass;
and $\tau_{\rm c}=20\Myr$ is the appropriate gas recycling time
(or `cloud lifetime').
We emphasise that this formula, derived on physical grounds,
results in a SN driving that is self-regulating,
with feedback from the evolving 3D structure;
the simulations evolve towards some equilibrium SN~II rate,
which becomes a testable output rather than a fixed input parameter.
(Although some calibration was performed to obtain an appropriate
value for $\tau_{\rm c}$;  see below.)
The location of the SNe is chosen from those positions satisfying
the criteria above,
with the probability at any position being proportional to the local density;
SN~II thus end up concentrated in the densest regions,
as is realistic.
Our model also retains an appropriate (small) number of Type~I SNe,
implemented with fixed temporal and spatial probabilities,
as in Korpi et al.\ (1999a);
these do not play a large role in the solution, however.

\section{Results and Discussion}

\begin{figure*}[tb]
\leftline{\includegraphics[width=2.25in]{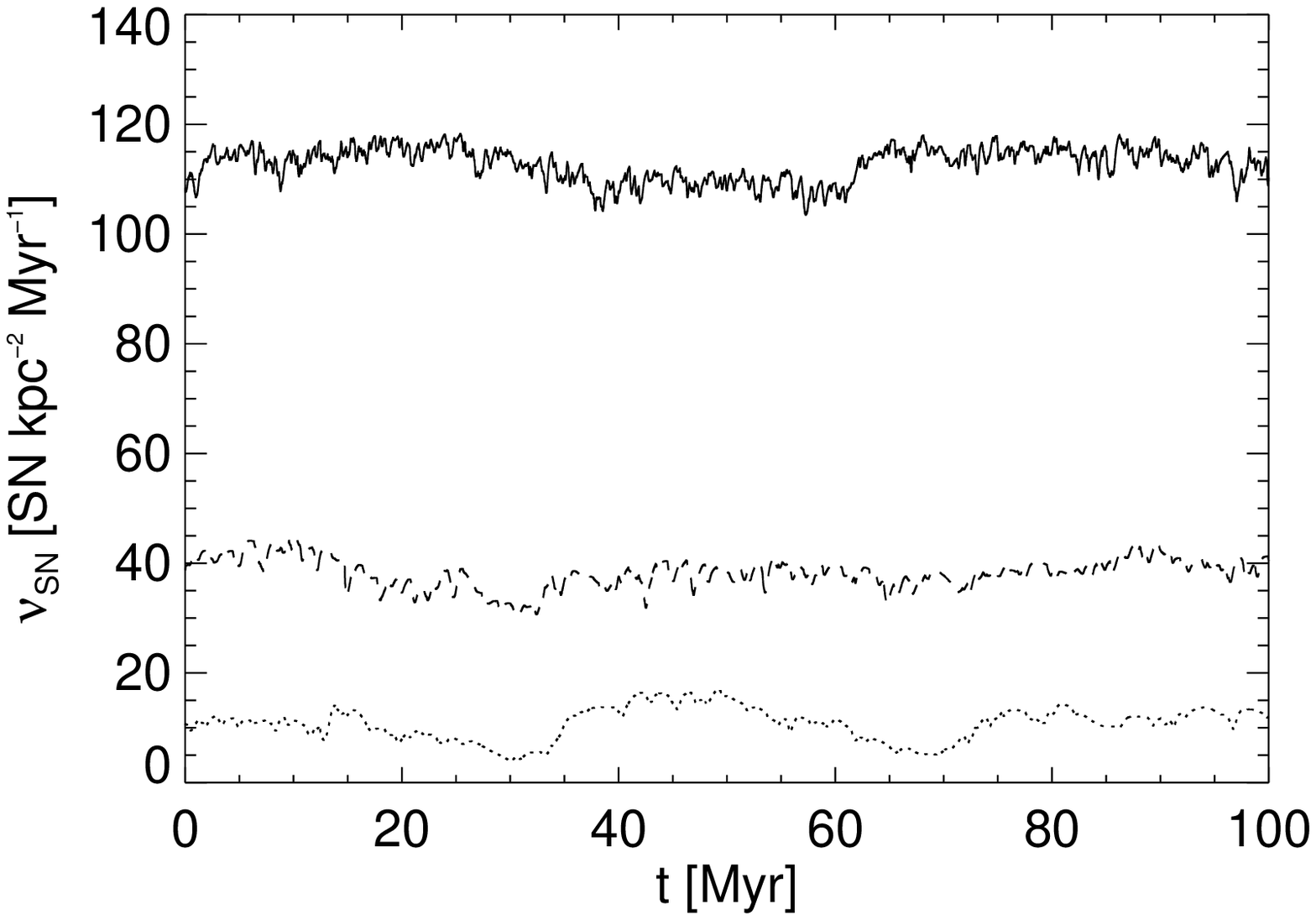}}
\vspace*{-41mm}
\rightline{\includegraphics[width=2.25in]{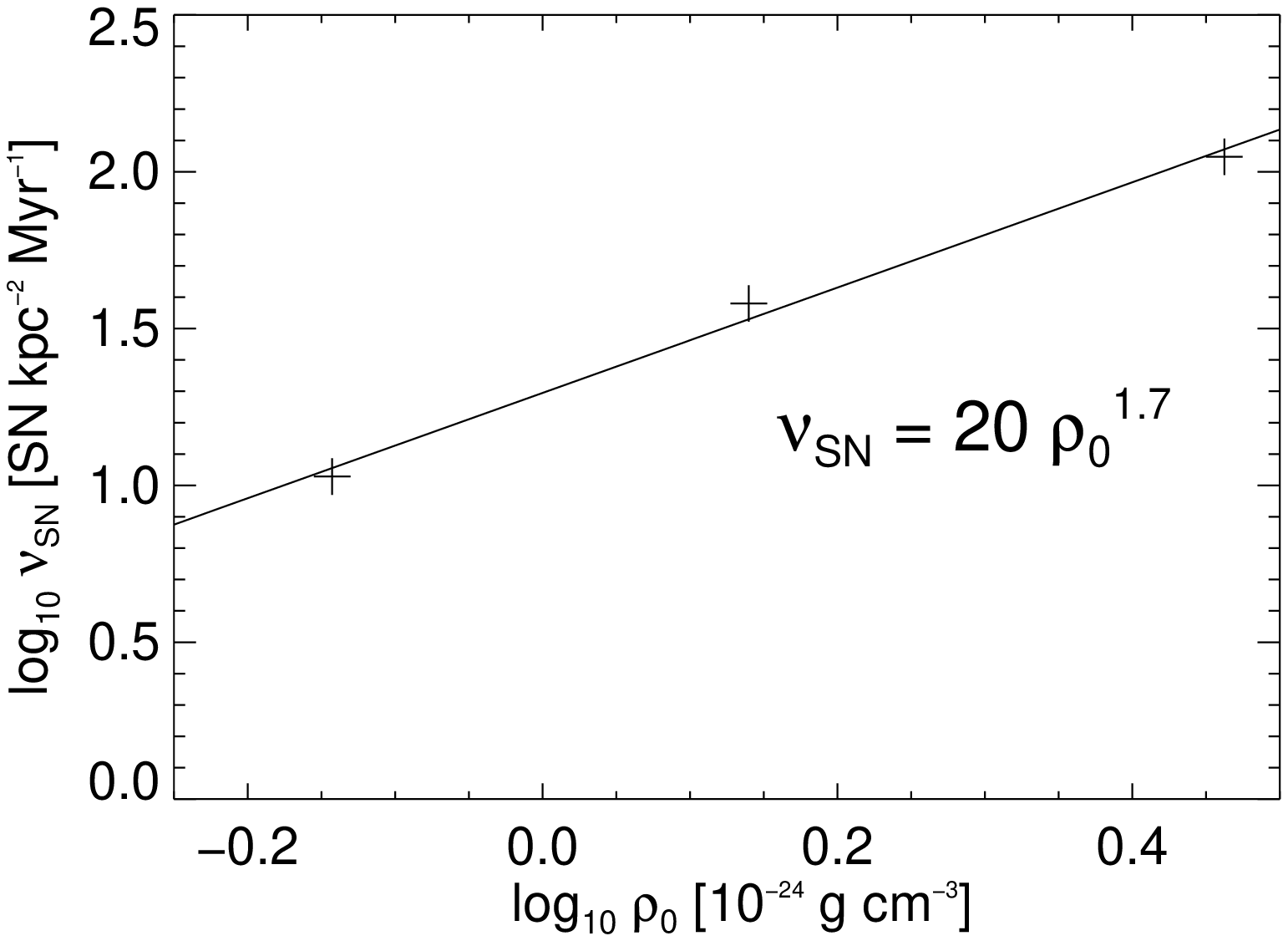}}
\caption{Time-evolution of the SN~II rate (left) 
for three simulations of differing density: 
the Arm (solid), Average (dashed) and Interarm (dotted) models
described in Table~\ref{tab:rho}.
The power law fit of the mean SN~II rates with density (right).}
\label{fig:time}
\end{figure*}

We have run simulations with three different mean densities,
chosen to model the ISM in regions within and between spiral arms,
as well as for an average value;
these are identified as the Arm (high $\rho$), Interarm (low $\rho$)
and Average cases, and their respective initial mid-plane densities, 
$\rho_{0}$, are given in Table~\ref{tab:rho}.
The runs were started in hydrostatic equilibrium,
using the density stratification profile of Ferri\`ere (1998),
but scaled by the density $\rho_{0}$.
After of order 20--40 Myr of evolution,
the system attains a convincing statistical equilibrium.
The time sequences shown in Figure~\ref{fig:time} show
the variations of the instantaneous SN~II frequency
after this initial period.

\begin{table*}[tb]                        \label{table1}
\begin{tabular}{lcccc}
\hline
                                        &Unit  &Interarm &Average & Arm \\
Initial mid-plane density,
        $\rho_{0}$ &$10^{-24}\gm\cmcube$ & 0.7   & 1.4   & 2.9   \\
\hline
SN II rate, $\nu\SN$ &${\rm kpc^{-2}}\,{\rm Myr}^{-1}$    & 11    & 38    & 111   \\
Hot gas filling factor, $f$     &                    & 0.12  & 0.07  & 0.04  \\
\hline
\end{tabular}
\caption{Three models with varying initial mid-plane density, $\rho_{0}$.
The filling factors are for gas with $T>10^{5}\K$, within $|z| < 0.2 \kpc$.
The initial magnetic field at the mid-plane is $B_{0} = 6 \muG$ in all cases.}
\label{tab:rho}
\end{table*}

Table~\ref{tab:rho} gives the average frequencies $\nu\SN$
arising from our SN implementation.
The value for the Average case,
$\nu\SNII\simeq40\kpc^{-2}\Myr^{-1}$
(which, naively extrapolated, gives $1/33 {\yr}^{-1}$ for the whole Galaxy),
was obtained after calibration of $\tau_{\rm c}$
to ensure just such a suitable average rate.
This value clearly cannot then be used in vindication of our model.
The rates for the other two simulations
subsequently arise with no further input,
however,
and the resultant variation is well fitted as
$\nu\SN=20 \kpc^{-2}\Myr^{-1} \,(\rho_{0}/\,10^{-24}\gm\cmcube)^{\;1.7}$
(see Figure~\ref{fig:time}).
We may consider our SN rate as a proxy for star formation rate (SFR),
and the exponent above is consistent with observed scalings 
of the latter quantity with gas surface density $\Sigma$:
$\; {\rm SFR}\propto\Sigma^\kappa$, with $\kappa=0.9$--1.7 (Kennicutt, 1998).
(For the density variations we impose,
$\Sigma$ scales directly with $\rho_{0}$.)
Whilst it is not clear that the two relations are directly comparable 
--- the SFR should depend on column-integrated density,
and our SN mechanism is more dependent on local volume density ---
this result remains of great interest, and potentially 
constitutes a meaningful test of the implementation of physics in our model.
Further work on this correlation is planned.

\begin{figure*}[tb]
\leftline{\includegraphics[width=2.25in]{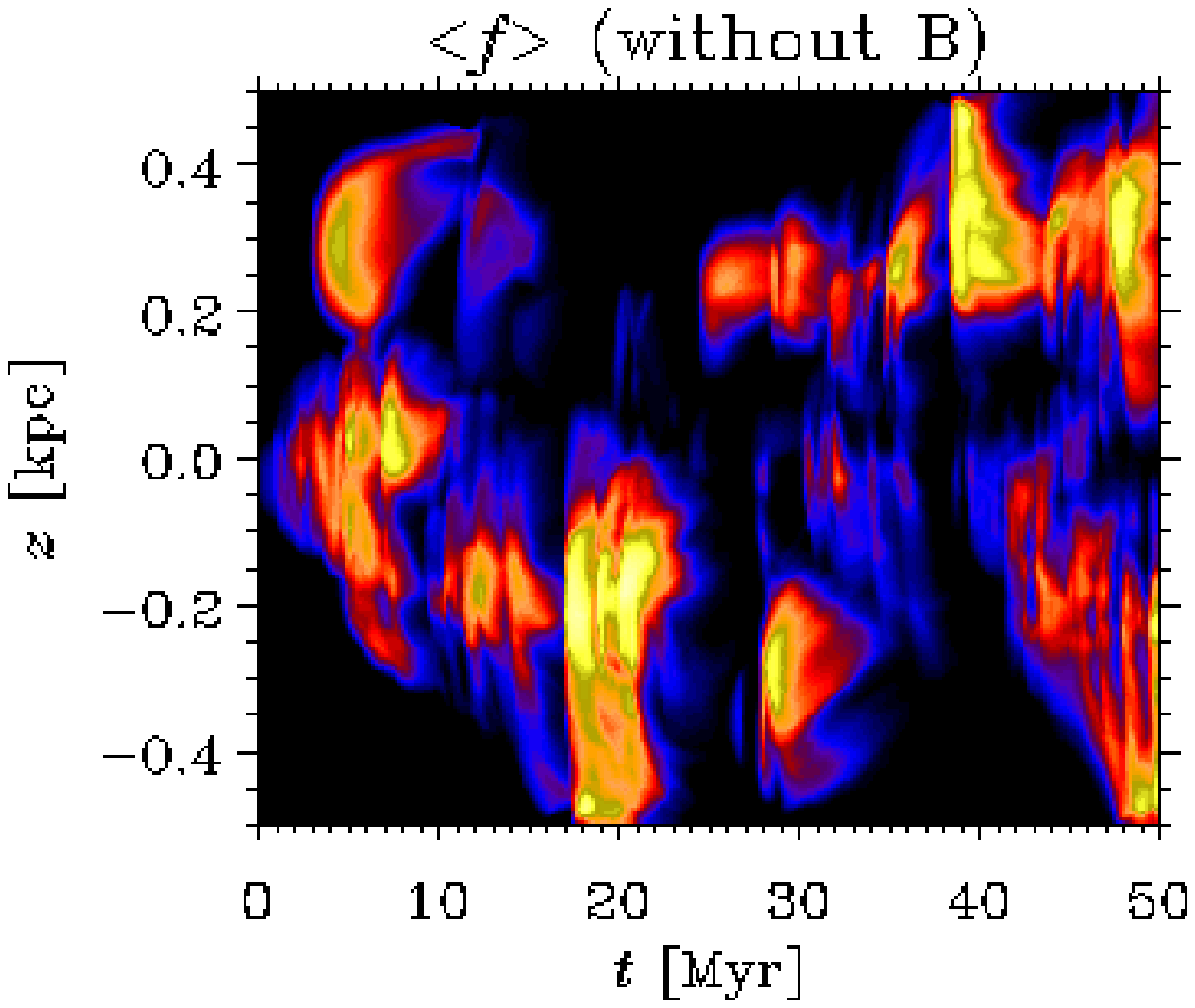}}
\vspace*{-46mm}
\rightline{\includegraphics[width=2.25in]{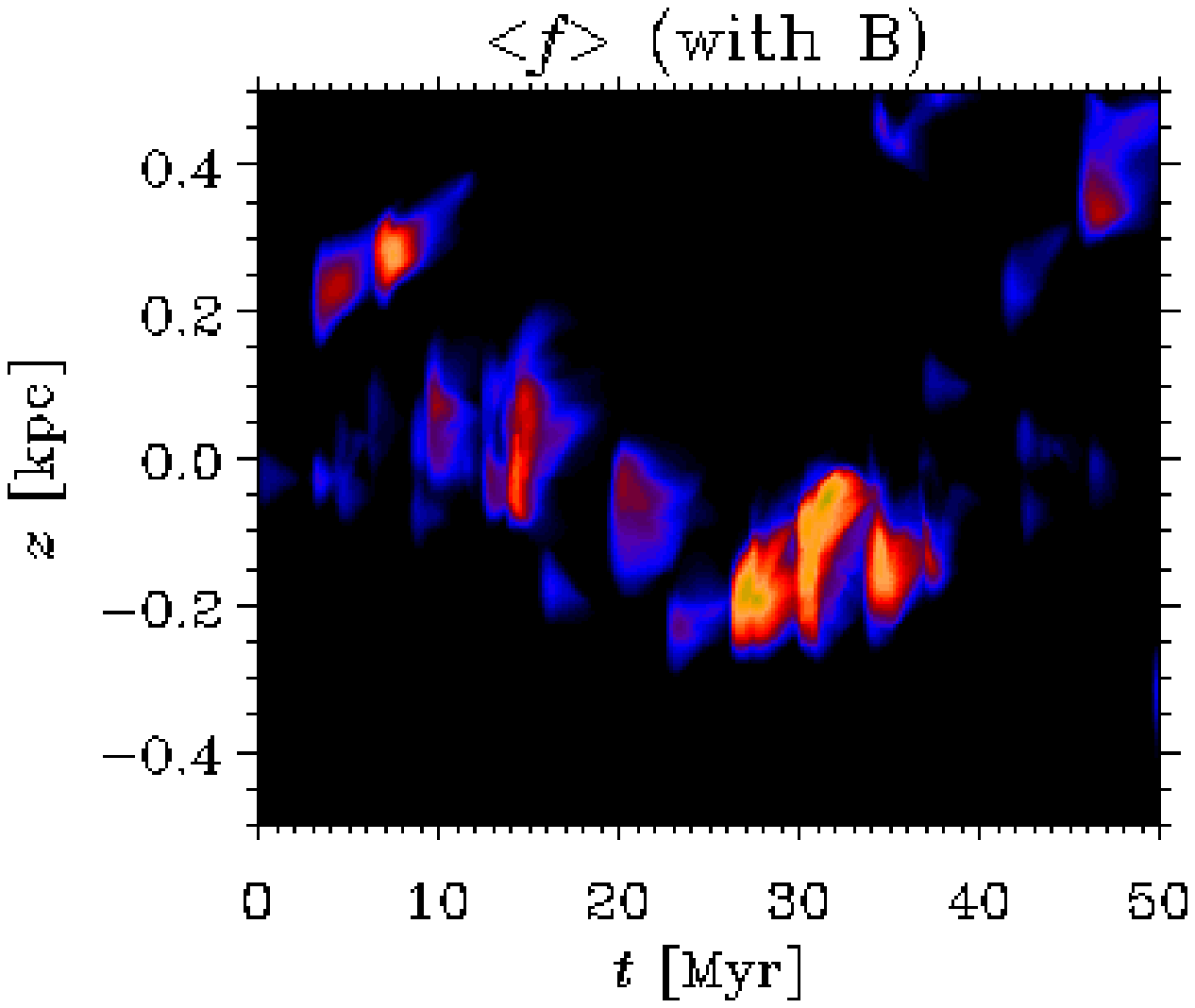}}
\caption{Hot gas filling factor $f$, as a function of height and time,
for runs with no magnetic field (left panel) and with $B_{0}=6\muG$ (right panel).
Filling factors range from 0 to 0.93;
higher values are indicated by lighter shades.
Both runs are for Interarm densities.}
\label{fig:ffm}
\end{figure*}

These simulations give volume filling factors of hot gas 
which are rather low (see Table~\ref{tab:rho});
plausible values for the Average case are of order 0.15--0.20.
This is a facet of the model that is particularly sensitive to
magnetic fields, however.
The runs described above had an initial azimuthal
field of the form $B_{y}=B_0\cosh^{-2}(z/0.3\kpc)$,
with $B_0=6\muG$.
This $B_{0}$ is a reasonable value for the {\em total\/} field
in the solar vicinity, but that field should be split
roughly equally between ordered and small-scale components.
A uniform field is extremely effective in confining
the expanding hot gas produced by SNe,
as is evident from Figure~\ref{fig:ffm}, which shows horizontally averaged
filling factors for two comparable runs with and without
such a field.
(We should comment that these filling factors may be 
rather sensitive to the precise definition of hot gas used, however.
Also our use of closed boundaries in $z$ may influence the vertical
structure of this quantity;  future work should remove this latter
limitation.)
The difference in mean filling factor is almost a factor of 2,
as can be seen in Table~\ref{tab:B}.
(A solution for an intermediate $B_{0}$,
not continued for so long,
gives consistent results.)
Re-considering the values in Table~\ref{tab:rho}
(obtained with the over-strong large-scale $B_{0}$)
in this light,
we believe that our model represents the evolution of such structures
in the gas in a reasonable way;
and we emphasise the clear importance of magnetic fields ---
neglected in most other simulations -- in this respect.

\begin{table*}[tb]                        \label{table2}
\begin{tabular}{lccc}
\hline
                                        &Unit  & No field & Strong field \\
Initial mid-plane field, $B_{0}$ & $\muG$ &  $0$  & $6$ \\ 
\hline
Hot gas filling factor &                       &  0.19  & 0.12 \\ 
\hline
\end{tabular}
\caption{Two models with varying initial mid-plane magnetic field, $B_{0}$.
Filling factors are defined as in Table~I.
The initial mid-plane density is $\rho_{0} = 0.7 \times 10^{-24}\gm\cmcube$
in both cases.}
\label{tab:B}
\end{table*}

\acknowledgements
Use of facilities at the Danish Center for Scientific Computing in Copenhagen,
and at UKAFF in Leicester,
is acknowledged.
This work was supported by the PPARC Grant PPA/G/S/2000/00528.

\theendnotes



\end{article}
\end{document}